\documentclass[fleqn,10pt]{wlscirep}
\usepackage[utf8]{inputenc}
\usepackage[T1]{fontenc}
\title{Hot-carrier dynamics in InAs/AlAsSb multiple-quantum wells}

\author[1]{Herath P. Piyathilaka}
\author[1]{Rishmali Sooriyagoda}
\author[2]{Hamidreza Esmaielpour}
\author[2]{Vincent R. Whiteside}
\author[2]{Tetsuya D. Mishima}
\author[2]{Michael B. Santos}
\author[2,+]{Ian R. Sellers}
\author[1,*]{Alan D. Bristow}

\affil[1]{Department of Physics and Astronomy, West Virginia University, Morgantown, WV 26506-6315, USA}
\affil[2]{Department of Physics and Astronomy, University of Oklahoma, Norman, OK 73019, USA}
\affil[+]{corresponding.sellers@ou.edu}
\affil[*]{corresponding.alan.bristow@mail.wvu.edu}

\begin{abstract}
A type-II InAs/AlAs$_{0.16}$Sb$_{0.84}$ multiple-quantum well sample is investigated for the photoexcited carrier dynamics as a function of excitation photon energy and lattice temperature. Time-resolved measurements are performed using a near-infrared pump pulse, with photon energies near to and above the band gap, probed with a terahertz probe pulse. The transient terahertz absorption is characterized by a multi-rise, multi-decay function that captures long-lived decay times and a metastable state of for an excess-photon energy of $>100$ meV. For sufficient excess-photon energy, excitation of the metastable state is followed by a transition to the long-lived states. Excitation dependence of the long-lived states map onto a near-direct band gap ($E{_g}$) density of states with an Urbach tail below $E{_g}$. As temperature increases, the long-lived decay times increase $<E{_g}$, due to the increased phonon interaction of the unintentional defect states, and by phonon stabilization of the hot carriers $>E{_g}$. Additionally, Auger (and/or trap-assisted Auger) scattering above the onset of the plateau may also contribute to longer hot-carrier lifetimes. Meanwhile, the initial decay component shows strong dependence on excitation energy and temperature, reflecting the complicated initial transfer of energy between valence-band and defect states, indicating methods to further prolong hot carriers for technological applications.
\end{abstract}
\begin{document}

\flushbottom
\maketitle

%\thispagestyle{empty}

%\noindent Please note: Abbreviations should be introduced at the first mention in the main text – no abbreviations lists. Suggested structure of main text (not enforced) is provided below.

\section*{Introduction}

Improving the light-to-electric conversion efficiency is crucial to the development of semiconductor photovoltaics. The detailed-balance limit for a single-junction silicon solar cells is dominated by the rapid cooling of charge carriers photoexcited above the semiconducting band gap, where their excess energy heats the lattice  \cite{shockley_detailed_1961}. Engineering hot-carrier solar cells (HCSC) was proposed to overcome this limit through extraction of the hot carriers before they cool, increasing the maximum theoretical efficiency from 33\% to 66\% under (unconcentrated) sunlight \cite{ross_efficiency_1982}. In order to do this, hot carriers must be extracted through energy-selective contacts faster than they emit optical phonons \cite{ferry_search_2019}. Since optical phonons subsequently decay into acoustic phonons \cite{klemens_anharmonic_1966,ridley_lo_1996}, engineering the photonic, electronic and phononic properties of semiconductors is important for improving HCSC devices \cite{basu_hot_2017,cebulla_hot_1988,esmaielpour_enhanced_2018,conibeer_hot_2014}.\par
    
For example, bandgap engineering of monolithic structures to include quantum confinement enhances optical absorption \cite{esaki_superlattice_1970}, type-II band-aligned quantum wells spatially separate electrons and holes to increase the excited-state carrier lifetime \cite{lee_investigation_2017-1, esmaielpour_enhanced_2018}, and structures with highly contrasting media can reduce the cooling through phonons \cite{ridley_electron-phonon_1982,shen_exciton-longitudinal-optical_1996}. These properties all occur in InAs/AlAs$_{0.16}$Sb$_{0.84}$ multiple-quantum wells (MQWs), where a hot-carrier distribution is shown along with extended carrier lifetimes as a result of inhibited phonon-phonon interactions \cite{esmaielpour_enhanced_2018}. In steady-state photoluminescence (PL), these MQWs have shown evidence of hot carriers \cite{tang_effects_2015}, non-monotonic emission energy as a function of lattice temperature due to the complicated valence band structure \cite{whiteside_valence_2019} and even intervalley scattering of electrons to the long-lived L-valley states \cite{whiteside_role_2019}. Transient absorption directly shows long-lived carriers complementing the evidence of hot carriers and the non-monotonic temperature dependence \cite{esmaielpour_enhanced_2018}. Phonon band-structure calculations supports suppression of Klemens’ process that successfully converts optical to acoustic phonons \cite{klemens_anharmonic_1966} in favor of the Ridley mechanism, which is less effective \cite{ridley_lo_1996}. Moreover, poor thermal conductivity due to the phonon impedance mismatch between the MQW layers results in a hot acoustic bath. Overall, the entire electron-to-thermal cooling pathway is slowed and optical phonon can be reabsorption by hot carriers, stabilizing them in non-thermalized states \cite{esmaielpour_enhanced_2018,zhang_explore_2020}. \par
    
Previously, transient-absorption measurements of these InAs/AlAsSb MQWs were performed at a single excitation energy above lowest interband transition, leaving unresolved the exact origin of the various competing phonon- and electron-scattering mechanisms that can contribute to the longevity of hot carriers \cite{esmaielpour_enhanced_2018,zhang_explore_2020}. In this paper, transient-absorption measurements are performed for a range of excitation conditions (charge-carrier density and excess-photon energies) to disambiguate the complexity of the charge carrier dynamics of InAs/AlAs$_{0.16}$Sb$_{0.84}$ MQWs for a range of lattice temperatures. Analysed results reveal contributions from the transitions from the various hole states to the electronic resonance in the InAs well. The most striking result is a strong plateau observed in the transient absorption for low excitation densities, sufficient excess-photon energy and at low-to-moderate lattice temperatures. 
    
\begin{figure}
\includegraphics{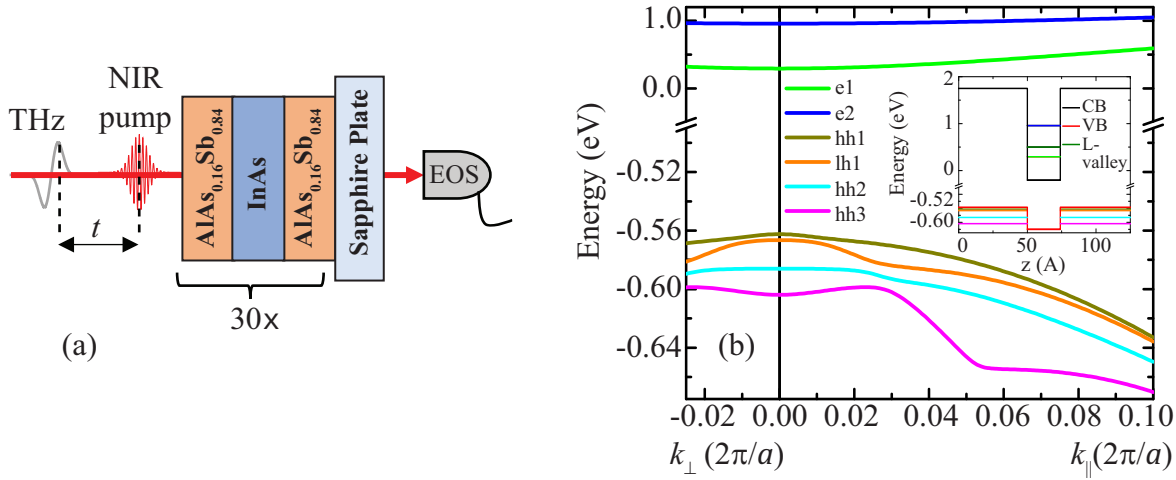}
\caption{\label{fig1}Schematic of the optical pump THz probe experimental geometry. EOS = electro-optic sampling. (b) Band structure of the InAs MQW superlattice. The inset shows the band minima and maxima of a single QW with confined energy levels at $k = 0$ cm$^{-1}$}
\end{figure}

\section*{Experimental}

Figure~\ref{fig1}(a) shows a schematic of the optical pump and terahertz (THz) probe spectroscopy performed on an InAs/AlAs$_{0.16}$Sb$_{0.84}$ MQW. The MQW structure is grown by molecular beam epitaxy and consists of 30 periods of 2.4 nm InAs wells and 10 nm AlAs$_{0.16}$Sb$_{0.84}$ barriers. A $\sim0.25$ cm${^2}$ chip is attached to a sapphire substrate using a transparent adhesive and was thinned chemo-mechanically to remove the GaAs substrate.\par

The experimental light source is a 1 kHz regenerative laser amplifier producing an $\sim$100 fs laser pulse with a center frequency of 800 nm. The laser emission is split into two replica pulses – one to generate and detect THz-probe pulses and another to generate the optical-pump pulses. Optical-pump pulses are generated in an optical parametric amplifier which has a signal tuning range of 1200 nm (1.03 eV) to 1600 nm (0.775 eV). Throughout the tuning range, the excitation density in the MQW is kept constant at $\sim10{^{13}}$ cm${^{-2}}$.\par

THz-probe pulses are generated by optical rectification in a 0.5 mm thick, (110)-cut CdSiP${_2}$ crystal by weakly focusing the 800 nm pulses at normal incidence and with linear polarization orientated along the $[1\Bar{1}0]$ axis to maximize the generation \cite{piyathilaka_terahertz_2019}. A high-density polyethylene low-pass filter transmits only the THz radiation which is collected and focused onto the sample using two off-axis parabolic mirrors (OAPMs). The sample is located in a cryostat with a controlled temperature range of 4 K to 300 K. The THz radiation transmitted through the sample is collected and refocused by two more OAPMs onto a 0.3 mm thick, (110) cut ZnTe electro-optic (EO) crystal. All components from THz source to detection crystal are enclosed in a light-tight enclosure that is purged with dry air to reduce absorption due to water vapor.\par 

An 800-nm pulse, split from the THz generation pulse, is time-of-flight controlled and used to detect the maximum THz-induced EO signal which is resolved by a Soleil-Babinet compensator, Wollaston polarizing prism and two balanced photodiodes. The EO signal is recorded with a lock-in amplifier as a function of pump-probe delay time (\emph{t}). The lock-in-amplifier responds at a modulation frequency used to mechanically chopping the pump path and which is synchronized to a sub-harmonic of the repetition rate of the laser amplifier.\par

Figure~\ref{fig1}(b) shows the calculated band structure for the MQW sample illustrating the in- and out-of-plane wavevector directions \cite{whiteside_valence_2019}. AlAs$_{0.16}$Sb$_{0.84}$/InAs/AlAs$_{0.16}$Sb$_{0.84}$ form quantum wells with type-II band alignment. They confine electrons with e1 and e2 conduction subband states in the InAs well, and heavy (hh1, hh2, hh3) and light holes (lh1) valence subband states in the AlAs$_{0.16}$Sb$_{0.84}$ barriers. The MQW bandgap is $E{_g}$($T=4$ K)= 0.857 eV arising from the e1-hh1 transition (denoted P${_1}$). Transition from all hole bands to e1 contributes to increased absorption above $E{_g}$, culminating in a peak absorption (P${_2}$) at the e1-hh3 transition energy near 0.89 eV. Not shown are the L-valley and X-valley local minima of the e1 conduction band\cite{whiteside_role_2019} which may also play a role in the scattering of the electron once photoexcited.\par

\begin{figure}
\includegraphics{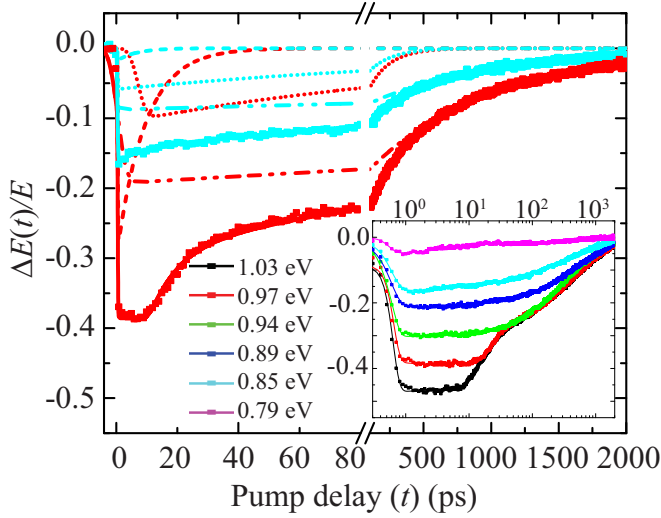}
\caption{\label{fig2}Transient differential THz-electric field [$\Delta{E(t)}\slash{E}$] through the MQW sample at 4 K for the excitation photon energies of 0.85 eV (cyan) and 0.97 eV (red). Multi-exponential fits are shown as solid lines, comprised of a fast (dash line), intermediate (dotted line) and slow (dash dot-dot line) component. The insets show a semi-log plot of the transient data and respective fits in the range 0.79 – 1.03 eV.}
\end{figure}

Temperature contraction of the P${_1}$ and P${_2}$ transition energies are taken into consideration when determining excess-photon energy $(\Delta(T)=E{_{pump}}-E{_g}(T))$ in this work. Hole state occupancy (due to alloy-intermixing defects in the well) and wavefunction localization/delocalization are also factored into account as the system transition from being quasi-type-I at 4 K to type-II at $\sim150$ K and quasi-type-II above that temperature \cite{esmaielpour_enhanced_2018}.

\section*{Results}

Figure~\ref{fig2} shows two examples of transient differential THz electric field $[\Delta{E(t)\slash{E}}]$ transmitted through the MQW sample which is cooled to 4 K and excited with 0.85 eV and 0.97 eV pump photon energies. The transients exhibit a fast rise with a multi-exponential decay that persist for over a nanosecond. The maximum amplitude of the transients increases with increasing pump photon energy and a plateau, lasting a few picoseconds in duration, emerges in the higher pump photon energy transients. This plateau is more clearly visible in the inset of the figure, where the pump-probe delay time is plotted on a logarithmic scale and data are shown for the excitation range of 0.79 eV to 1.03 eV. This excitation range corresponds to an excess-photon energy range of -50 meV$<\Delta<$175 meV at 4 K and blue shifts with increasing temperature. Transients are also recorded for 100 K, 200 K and 300 K.\par

To capture the plateau, the transients are fit to a three-component solution to the canonical rate equation,
 $\Delta{E(t)}\slash{E}=\sum_{i=1}^{3}A{_i}\{\frac{1}{2}\text{erfc}[-(\Delta{t})\slash\tau{_{Ri}}]\}  \exp[-(\Delta{t})\slash\tau{_{Di}}]$, where $\Delta{t}=t-t{_0}$, and A${_i}$, $\tau{_{Ri}}$, and $\tau{_{Di}}$ are amplitude, rise time and decay time for the ith component respectively. For best fitting, the first rise component ($\tau{_{Ri}}$) is chosen to be the cross-correlation of the pump and probe pulses, rounded to $\sim200$ fs. The second and third rise components ($\tau{_{R2}}$ and $\tau{_{R3}}=\tau{_{R2}}$) are free parameters to best fit the plateau region at early delay times in the transients. The various parameters extracted from the fits are plotted in the next few figures along with temperature- and energy-dependent analysis.\par
 
\begin{figure}
\includegraphics{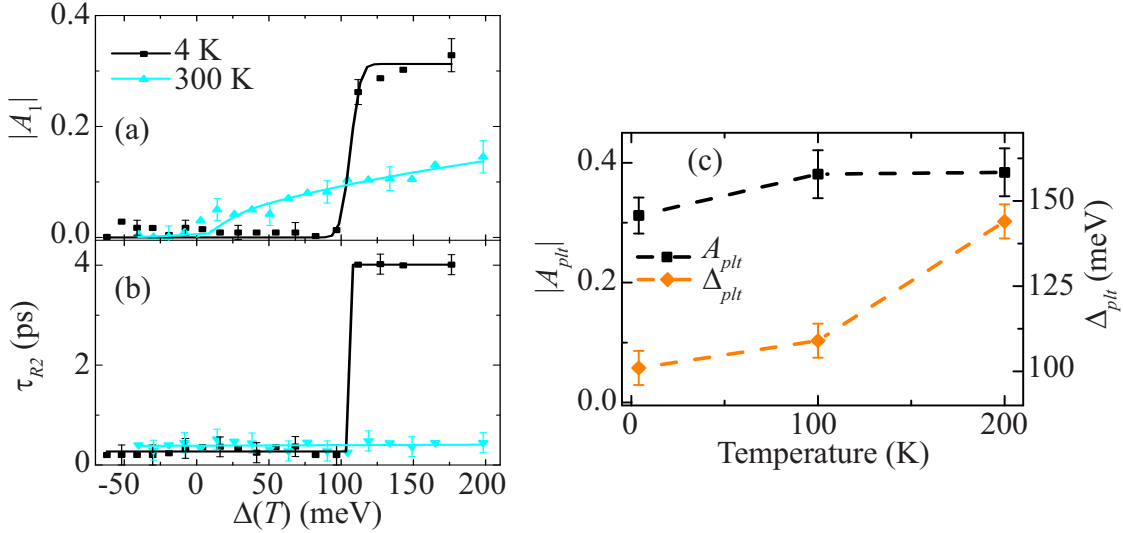}
\caption{\label{fig3}Excess-photon energy dependent of (a) $|A_{1}|$ and (b) $\tau_{R2}$ from the fitted transients for 4 K and 300 K, revealing a plateau in the former. (c) Temperature dependence (up to 200 K) of the on magnitudes for fast decay mechanism black square ($|A_{plt}|$) for step function and orange diamond  for energy onset of plateau}
\end{figure}

Figure~\ref{fig3}(a) and (b) show extracted magnitude ($|A{_1}|$) and rise time ($\tau{_{R2}}$), as a function of $\Delta (T)$ for $T= 4$ K and 300 K. At 4 K, both $|A{_1}|$ and the $\tau{_{R2}}$ are correlated, exhibiting a transition at $\Delta{_{plt}}\approx$ 100 meV. $\Delta{_{plt}}$ corresponds to the onset of the plateau in the transients and is best understood by comparing the constituent transients in Figure~\ref{fig2}. For 0.85 eV excitation ($<\Delta{_{plt}}$), $|A{_1}|$ is small (exhibiting a weak free-carrier absorption signal) and $\tau{_{R2}}\approx\tau{_{R1}}$, indicating excitation into long-lived states that are responsible for strong PL \cite{tang_effects_2015}. By contrast for 0.97 eV excitation ($>\Delta{_{plt}}$), $|A{_1}|$ is significant, exhibiting both a fast rise and decay. The fast decay is simultaneous with a slow rise of the longer-lived states, indicating a subband transfer of charge carriers.\par

The magnitude of the metastable plateau is modelled as $\left|A_{1}(\Delta, T)\right|=A_{p l t}(T)\left(\frac{1}{2} \operatorname{erfc}\left\{-\left[\Delta-\Delta_{p l t}(T)\right] / w\right\}\right)$, where $A_{plt}$(T) is the temperature-dependent amplitude of the plateau contribution and $w$ is the transition width (limited by the spacing of the tuning of the laser’s center photon energy). A similar expression can be determined for $\tau_{R2}$ with an addition of a $\tau_{R1}$-like term below $\Delta{_{plt}}(T)$. Figure~\ref{fig3}(c) shows that $A{_{plt}}(T)$ increases up to T= 200 K as the band structure contracts and the density of states at each energy interval increases. Figure~\ref{fig3}(c) shows that $\Delta{_{plt}}$ also increases up to 200 K, most likely due to thermal-expansion-induced strain (resulting from the sample being adhered to the sapphire substrate that modifies the band positions with respect to one another \cite{wilmer_role_2016}.)\par

\begin{figure}
\includegraphics{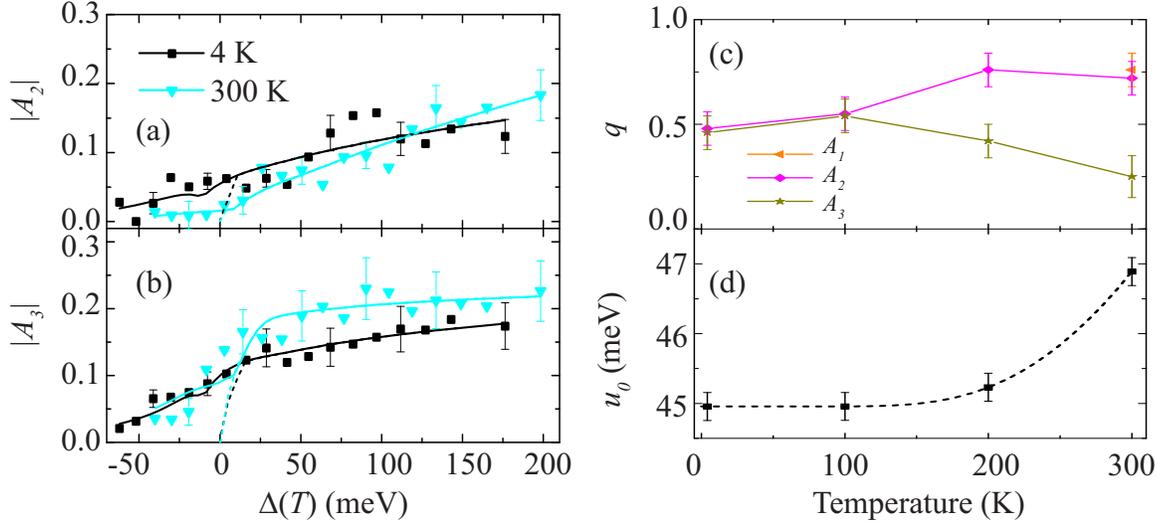}
\caption{\label{fig4}Excess-photon energy dependence of (a) $|A_{2}|$ and (b) $|A_{3}|$. Temperature dependence of (c) power value, \emph{q}, to determine the band-to-band transition type and (d) Urbach energy, $u_{0}$, for below-gap thermal and non-thermal contributions.}
\end{figure}

In contrast to the 4 K data, at 300 K the $\Delta$-dependence of $|A_{1}|$ does not show the plateau behavior. Instead $|A_{1}(\Delta)|$ comprises of a weak exponential growth below $\Delta=0$ and a saturating power-law growth above $\Delta=0$. This behavior is also seen for the magnitudes of the two slower decay components $|A_{2,3}|$; see Figure~\ref{fig4}(a) and (b). Above $E_{g}$, the magnitude is indicative of the interband absorption and $\propto|A_{(i,q)}|\Delta^q$, where \emph{q} is determined by the density of states \cite{oppo_optical_2020}. [Namely, $q=0$ for localized states, $q=0.5$ for purely direct-gap transitions in bulk, $q=1$ for transitions in an infinite quantum well and $q=2$ for purely indirect transitions in bulk.] Figure~\ref{fig4}(c) shows that $q\approx0.5$ at low temperature for all three constituents of the transient, indicating that the absorption is direct and bulk-like, which is unsurprising for the 30-periods InAs/AlAs$_{0.16}$Sb$_{0.84}$ MQW with fairly large penetration depth of the wavefunctions into the barriers and with alloy-intermixing states at the well interface \cite{bansal_alloying_2007}. As temperature increases, the fast and intermediate components increase by about 40\% (and overlaps with the \emph{q} value extracted from $|A_{1}|$ when the plateau vanishes at room temperature), while \emph{q} decreases by about 50\% for the slow component. A weighted sum may be able to determine the fraction of direct, indirect, allowed and forbidden transition contributions \cite{bristow_two-photon_2007}.\par

Below $E_{g}$, the magnitude of the transients shows an absorption tail $\propto{|A_{(i,u)}|}\exp(\Delta\slash{u_{0}})$, where the Urbach energy $(u_{0})$ characterizes the extent of the absorption tail into defect states and other disorder \cite{urbach_long-wavelength_1953}. Figure~\ref{fig4}(d) shows the temperature dependence of the Urbach energy exhibited in all transient component. At low temperature, $u_{0}\approx45$ meV and it increased by $\sim4$\% at 300 K. Lattice vibrations and dynamic structural disorder both effect $u_{0}(T)$ \cite{mondal_localization_2017}. The former only increases $u_{0}(T)$ as a function of \emph{T} due to a growing phonon distribution \cite{seyf_rethinking_2017}, whereas dynamic structural disorder can reduce $u_{0}(T)$ and has been related to the onset of medium-range order, typically in amorphous materials. For the high-quality MQW sample, the lattice vibration contribution to $u_{0}(T)$ must dominate. This is supported by analysis of Urbach energy using the Einstein model, $u_{0}(T)=(k_{B}\Theta)\slash\sigma_{0}[(1+X)\slash2+1\slash(\exp(\Theta\slash{T})-1)]$, where $X$ is a dimensionless parameter related to structural disorders, $1\slash\sigma_{0}$  is a dimensionless constant related to electron-phonon coupling and $\Theta$ is the Einstein temperature (which is $\sim\frac{3}{4}$ of the Debye temperature). Here, $X=4.29\times10^{-3}$, approaching the value for a perfect crystal ($X=0$)\cite{cody_disorder_1981}, due to very mild disorders \cite{rhiger_infrared_2020,gozu_interface_2006,shen_tamm_1995}.\par

\begin{figure}
\includegraphics{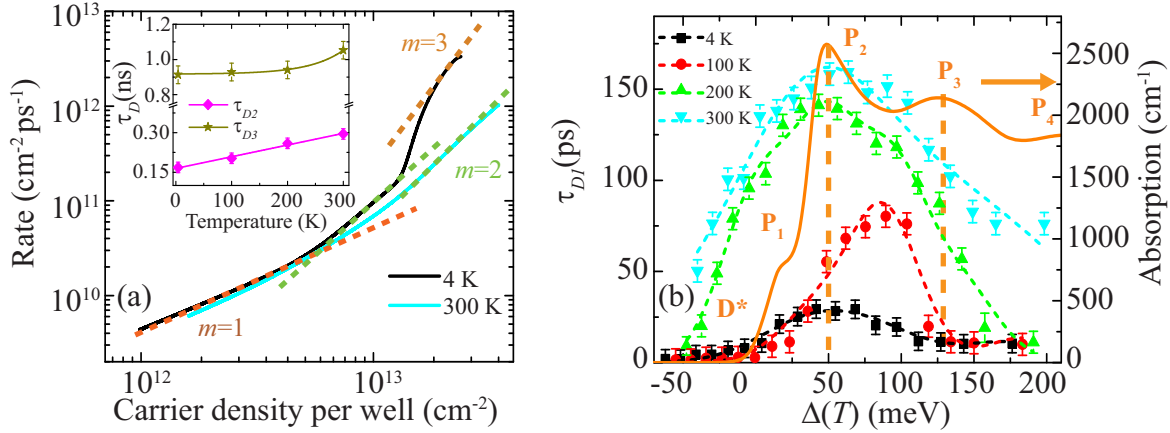}
\caption{\label{fig5}(a) Relaxation and/or recombination rate of excited-state carriers for 1.03-eV excitation at 4 K (black) and 300 K (cyan). Dashed guides to the eye represent known recombination mechanisms. Inset: temperature dependence of decay times $\tau_{D2}$ and $\tau_{D3}$, averaged over excess-photon energy. (b) Excess-photon energy dependence of fast decay time ($\tau_{D1}$) fit with five Gaussian distributions centered at D* and P$_{1}$-P$_{4}$ (dashed lines). The latter are determined from the simulated optical absorption (solid line).}
\end{figure}

The long-lived components are responsible for the strong PL and their decay times appear to be somewhat unaffected by the pump photon energy. For 4 K and averaged over the pump detuning, $\tau_{D2}=0.16$ ns and $\tau_{D3}=0.91$ ns; see inset of Figure~\ref{fig5}(a). These values both increase with increasing temperature, which is associated with the increase of the phonon distribution and the expected phonon-induced stabilization of the hot carriers \cite{esmaielpour_enhanced_2018}.\par

Decay mechanisms can be determined from recapturing the rate-equation by inverting the transient $|\Delta E(t)\slash E|^{2}$ and converting it into $\partial[n,p]\slash\partial t$ versus $[n,p]$ by use of the linear absorption coefficient for excitation pulse. In this case, \emph{n} and \emph{p} are the photoexcited electron and hole populations \cite{senty_inverting_2015} and the slope indicates the decay mechanism. Figure~\ref{fig5}(a) shows the rate equation for 1.03 eV excitation ($\Delta>\Delta_{plt}$) at 4 K and 300 K on a logarithmic scale. Overlaid are guides to the eye demarcating the $[n,p]^{m}\slash\tau$ slopes with $\tau$ as the instantaneous decay time. In this analysis, $m=1$ corresponds to Shockley-Read-Hall (SRH) dynamics, which involve recombination of electrons or holes with defect state (hence, $\sim{n}\slash\tau$ or $\sim{p}\slash\tau$); $m=2$ corresponds to interband radiative recombination ($\sim{np}\slash\tau$) or trap-assisted Auger scattering, where SRH recombination results in scattering of a carrier of the same species ($\sim{n^{2}}\slash\tau$ or $\sim{p^{2}}\slash\tau$); and $m=3$ corresponds to Auger scattering, where interband recombination results in scattering of a carrier ($\sim{n^{2}}p\slash\tau$ or $\sim{np^{2}}\slash\tau$) \cite{takeshima_auger_1972,metzger_auger_2001,marko_role_2003}.\par 

For 4 K, the slope indicates the dominance of Auger recombination at high carrier concentration as might be expected in a high-confinement nanostructure. This corresponds to the plateau region at the beginning of the transient. Auger recombination gives way to trap-assisted Auger or radiative recombination and finally SRH dynamics as the carrier concentration continually decreases. In contrast, 300 K data does not show the initial Auger scattering, $m=2$ processes persist to much higher carrier concentrations and the trend is otherwise similar to the low temperature results at lower carrier concentrations. The temperature behavior is consistent with PL results \cite{tang_effects_2015} only if (i) Auger-scattering during the plateau region leads to a persistence of hot carriers followed by a strong burst of radiative emission as the carrier concentration decays and (ii) radiative recombination dominates over trap-assisted Auger when excitation does not lead to the plateau behavior. For this $\Delta$, conversion from dynamics that include the plateau response at low temperature to dynamics without it at near-room temperature is likely related to the higher-temperature delocalization of the valence bands throughout the MQW structure \cite{tang_effects_2015}.\par 

The high carrier-excitation regime varies significantly with temperature; therefore, it is reasonable to expect that – unlike the long-lived decay times – $\tau_{D1}$ exhibits strong dependence on both $\Delta$ and $T$; see Figure~\ref{fig5}(b). Data are fit with five Gaussian distributions each with a center position that corresponds to either the defects D* (below $E_{g}$) or P$_{1}$ through P$_{4}$ and with widths that are commensurate with the Urbach tail or the width of the calculated optical absorption also shown in the figure \cite{whiteside_valence_2019}. At low temperature, $\tau_{D1}$ increases from just below $E_{g}$ to a maximum at the P$_{2}$ energy identified in absorption calculations and decreases above that energy. At 100 K, the maximum of $\tau_{D1}$ shifts between P$_{2}$ and P$_{3}$ and is about three times slower. Excitations at the band edge (P$_{1}$) now have a faster decay, while excitation at or above P$_{2}$ are stabilized by phonons and the Auger recombination process. This result clearly indicates that the various hole states indeed have different decay times. At 200 K, the MQW band alignment is known to be type-II and $\tau_{D1}$ is radically increased across the entire pump detuning range, exhibiting slower decays from defect contributions below P$_{1}$ and ever slower decays from states above. This behavior is similar for 300 K with a further slowing at the higher range of $\Delta$ where it is presumed the increased phonon distribution better stabilizes the hot carriers. For all temperatures, large $\tau_{D1}$ values are limited at an upper energy of P$_{3}$ which corresponds well with $\Delta_{plt}(T)$ determined from the amplitude analysis. Above this upper energy limit all excitations are fast and result in transitions to the longer-lived states. Only the data for 300 K show a significant $\tau_{D1}$ above P$_{3}$, where the P$_{4}$ resonance also contribute to slowing the initial decay. %\par

\section*{Discussion}

The energy and temperature dependence of the charge-carrier dynamics in the InAs/AlAs$_{0.16}$Sb$_{0.84}$ MQW are complicated due to the multiple hole subbands, the unusual interaction with phonons, the changing of the localization of the holes with temperature and the possible interaction with moderate numbers of defects that give a non-zero contribution to the dynamics below the band gap. Analysis of the dynamics encourage several questions, the most prominent being \emph{what is the origin of the metastability at early times and the resulting plateau seen in various components of the deconvolved transients?}\par 

One possible origin is the excitation at $k\neq0$ states, where due to valence-band mixing there are saddle-points and even local minimum in hh3 such that, the dispersion that can slow the relaxation of the holes and limit recombination with e1 electrons. In the simulations, the exact energy of the local minima is more error-prone than determination of $k\neq0$ states. Reasonably errors of $\sim10-20$ meV provide a sufficient margin to suppose that optical excitation from hh$3\rightarrow${e1} occurs at $\Delta\approx100$ meV at low temperature and slightly more at elevated temperatures. If this argument is upheld, it may also explain why there are so many components observed in the transients, since the signal would be dominated by the hole bands, both in terms of the response to excitation and in terms of the mechanisms of decay. For example, it is clear that the metastable state at early times does eventually relax into long-lived states where stable hot carriers persist through the inhibited electron-phonon interaction. This is consistent with increasing temperature delocalizing the hole wavefunctions and allowing this metastability to no longer play a role in the dynamics.\par 

Alternatively, even though the onset of the plateau occurs several 100’s of meV below the excitation energy required to expect easy inter-valley scattering to the L-valley  \cite{ferry_search_2019,whiteside_role_2019,zhang_explore_2020,hess_intervalley_1996,esmaielpour_exploiting_2020}, the electric field of the THz probe is approximately double the required field to result in a Gunn effect in these samples $\sim$17 kV/cm \cite{whiteside_role_2019}. However, if hot electrons persist and are even heated through electron phonon interactions after their initial excitation some degree of band tilting due to the THz field may scatter e1 electrons into the L-valley \cite{ferry_challenges_2020} to contribute to the complicated transient response. This affect does not quite match with initial experiments performed for a range of probe electric field strengths (straddling $\sim$17 kV/cm), where it might be expected that at lower field strengths the sharp edge of the plateau in $|A_{1}|(\Delta)$ may soften or the plateau may even vanish. Even at a probe field strength of $\sim$5 kV/cm the onset of the plateau is still sharp at an unmodified $\Delta_{plt}$.\par 

\begin{figure*}[ht!]
\includegraphics{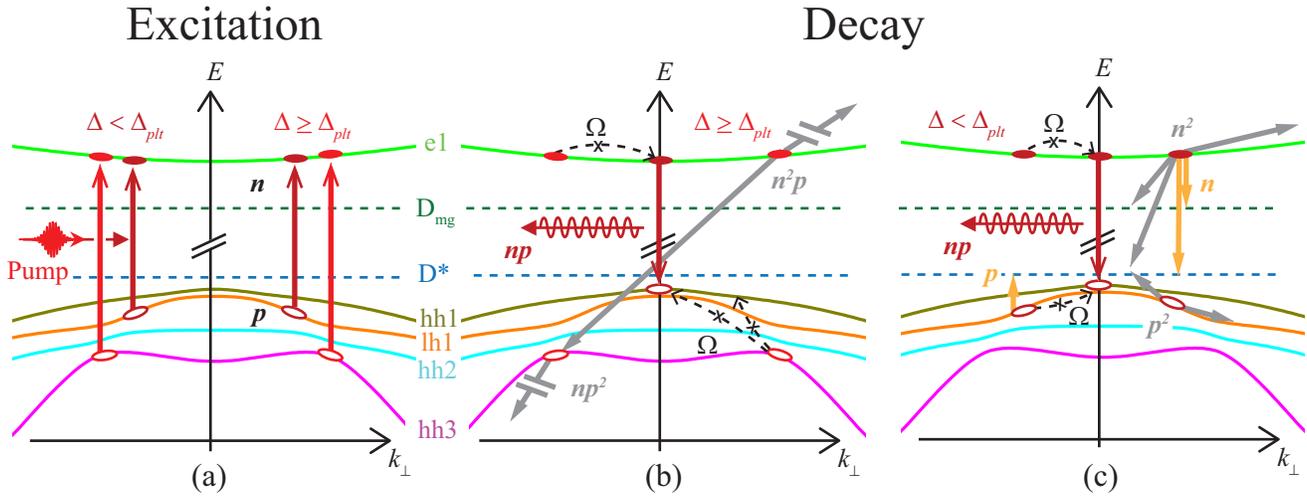}
\caption{\label{fig6} (a) Interband Excitation by the optical pump creates holes (\emph{p}) and electron (\emph{n}) in the valence and conduction bands at non-zero in-plane wavevectors. (b) Excitation with pulse detuning at $\Delta\geq\Delta_{plt}$ results in decay by Auger-scattering $[\propto{n^{2}}p,np^{2}]$, followed by intraband relaxation via phonon emission [$\Omega$], then (c) trap-assisted Auger $[\propto{n^{2}},p^{2}]$ and radiative recombination $[\propto{np}]$ and finally by Shockley-Read-Hall dynamics $[\propto{n,p}]$. Excitation with $\Delta<\Delta_{plt}$ results in decay only through the stages outlined in (c).}
\end{figure*}

A final and less likely alternative is the trapping of charge carriers in defect states. Despite the high-quality epitaxial growth InAs/AlAsSb quantum wells are known to have disorder  \cite{gozu_interface_2006,kroemer_61a_2004,dabrowski_isolated_1989} – as evidenced by the Urbach tail. There are a host of potential unintentional defects that may trap the carriers, but without testing the specific sample with electron paramagnetic resonance spectroscopy it would be hard to identify a particular defect that may be expected to result in long-lived carrier trapping.\par 

Based on this discussion, Figure~\ref{fig6} summarizes the non-equilibrium dynamics of photoexcited electrons and holes in the MQWs. For excess photon-energies below the onset of the metastable plateau in the dynamics ($\Delta<\Delta_{plt}$), the excitation amplitude shows a mostly square-root dependence as a function of $\Delta$. This is indicative of exciting direct transitions hh1-e1, lh1-e1 and hh2-e1, with carriers promoted symmetrically into parabolic band at in-plane wavevector $k_{\perp}>0$. Carriers then undergo charge separation into type-II-aligned wells and barriers, trap-assisted Auger scattering ($\propto{n^{2}},p^{2}$) with InAs interface traps \cite{tang_effects_2015,shen_remote_1993,cesari_role_2009}, radiative recombination ($np$) responsible for PL \cite{tang_effects_2015} and SRH non-radiative recombination ($\propto{n,p}$) with interface and mid-gap states \cite{tang_effects_2015}. Weak electron-phonon coupling and phonon-phonon scattering \cite{garg_phonon_2020} slows carrier thermalization and intraband relaxation, instead stabilizing the hot carriers and resulting in prolonged decay times. In addition, for $\Delta>\Delta_{plt}$ at temperatures $\geq200$ K, hh3-e1 transitions are symmetrically excited at $k_{\perp}>0$, where local minima in the hh3 states are suspected to further stabilize the holes and even reduce charge separation by reducing wavefunction overlap. During the plateau regime, scattering to $k=0$ states is further reduced and conventional Auger scattering ($\propto{n^{2}}p,np^{2}$) occurs pushing carriers deep into their respective bands. In InAs, Auger scattering is dominant for holes \cite{takeshima_auger_1972,metzger_auger_2001,marko_role_2003}. Slowly, the carrier density decreases via intra-valence band scattering and the dynamics revert to that seen for excitation $\Delta>\Delta_{plt}$.

\section*{Conclusion}

In conclusion, excited metastable carriers can be observed when the excess photon energy exceeds $\sim$100 meV at sufficiently low lattice temperature. This metastability most likely corresponds to local minima in the non-zero in-plane wavevector states of the hh3 band and leading to an accumulation of carriers that have a slow intra-valence cooling. The dynamics of the meta-stable states are dominated by Auger scattering that can potentially increased hot-carrier extraction, assuming that extraction can be achieved faster than the cooling. Once intra-valence band scattering occurs the electron and hole recombination rates are typically slow due to charge-carrier separation in the MQW structure and a slow cooling mechanism due to phonon-phonon coupling limits and a thermal conductivity mismatch in the structure. Hence, the long-lived carrier lifetimes that are already suitable for hot-carrier extraction in solar cells could be further enhanced by making use of the metastable state at early times.\par

Additionally, the complicated optical response of devices similar to these InAs-based MQW systems is well suited to investigation of the dynamics through both multi-exponential and inversion analysis. This approach complements device transport measurements and conventional photophysics approaches, such as photoluminescence, by providing detailed understanding of the dynamics mechanisms and associated states and should be considered as an integral part of designing hot-carrier Solar-cell devices.%\par

\bibliography{References}

\section*{Acknowledgements}

This research is funded through Department of Energy EPSCoR Program and the Office of Basic Energy Sciences, Material Science and Energy division under Award No.$\#$DE-SC0019384.

\section*{Author contributions statement}

H.P.P., R.S. and A.D.B. performed optical experiments and analysis. H.E., V.R.W., T.D.M., M.B.S. and I.R.S. performed sample growth, preparation and calculations. All authors wrote the manuscript.

\section*{Additional information}

%To include, in this order: 
%\textbf{Accession codes} (where applicable);

\textbf{Competing interests} The authors declare no competing interests. 

The corresponding author is responsible for submitting a \href{http://www.nature.com/srep/policies/index.html#competing}{competing interests statement} on behalf of all authors of the paper. 
%This statement must be included in the submitted article file.
\end{document}